\documentclass[epjST]{svjour}
\usepackage{graphics}
\begin{document}
\title{A study on dynamical complexity of noise induced blood flow}
\author{Bo Yan \inst{1} \and Sayan Mukherjee\inst{2} \fnmsep\thanks{\email{msayan80@gmail.com}} \and Shaobo He\inst{3}}
\institute{School of Computer Science and Technology, Hunan University of Arts and Science, Changde 415000, China.
\and Department of Mathematics, Sivanath Sastri College, Kolkata 700029, India.
\and School of Physics and Electronics, Central South University, Changsha 410083, China.}
\abstract{In this article, the dynamics and complexity of a noise induced blood flow system have been investigated. Changes in the dynamics have been recognized by measuring the periodicity over significant parameters. Chaotic as well as non-chaotic regimes have also been classified. Further, dynamical complexity has been studied by phase space based weighted entropy. Numerical results show a strong correlation between the dynamics and complexity of the noise induced system. The correlation has been confirmed by a cross-correlation analysis.} 
\maketitle
\section{Introduction}
\label{intro}
In a human body, the heart controls the blood flow through various blood vessels. Based on the functions, the blood vessels are classified by arteries, capillaries and veins. Arteries carry fresh oxygen from the heart to all of the body's tissues. A capillary is a thin vessel, connect the arteries and the veins. On the other hand, blood contains less oxygen and rich carbon dioxide that back to the heart through a vessel. Abnormality in arteries affects the heart to pump sufficient blood that needs the human organs. As a result, the aforesaid cyclic process faces an irregularity. The diameter of the vessel ($D$) and pressure of the blood ($P$) takes an important role in the blood flow. \par
To predict the blood flow phenomenon, several models have been proposed in \cite{RefA1,RefA2,RefA3,RefA4,RefA5,RefA6,RefA7}. Nonlinear dynamics \cite{RefA8,RefA9,RefB1,RefB2,RefB3,RefB4} is an efficient one, which can predict the long-term dynamics of a blood flow. Different types of long-term features, viz; stability, chaos, hyperchaos, etc have been proposed to identify the regular and irregular dynamics in a system \cite{RefA8,RefA9,RefB1,RefB2,RefB3,RefB4}. Irregular dynamics can always be observed in a complex system and recognizes by either chaos or hyperchaos, at least in, deterministic sense \cite{RefA9,RefB1,RefB2}. On the other hand, regular behaviour of a system can be recognized by stability analysis \cite{RefA8,RefB3,RefB4}. In \cite{RefA7}, authors have proposed a nonlinear model consisting of $D$ and $P$ that can only describe the stable behaviour of the blood flow only. Further, chaotic phenomena have also been observed using periodic disturbance \cite{RefB5,RefB6}. It implies that external disturbance can increase intricacy in dynamics. Various researchers have established the existence of the noise induced complex behaviour in different systems \cite{RefC1,RefC2,RefC3,RefC4,RefC5}. Indeed, the chaotic dynamics of noise induced blood flow and its complexity have not studied.\par
In general, chaos can be identified by measuring exponential divergence of the phase space trajectory. Method of Lyapunov exponent is one of the powerful tools that can measure the exponential divergence properly \cite{RefD1,RefD2,RefD3}. For an unknown mathematical model, the method of phase space reconstruction \cite{RefD4,RefD5,RefD6,RefD7} does not reveal a proper result which is discussed in \cite{RefD8,RefD9} [28,29]. In \cite{RefE1}, a method has been proposed that can successfully distinguish chaotic as well as regular dynamics in a system \cite{RefE1,RefE2}. Moreover, it does not require phase space reconstruction \cite{RefE1}. The invariant nature of a chaotic phase space can be characterized by dynamical complexity of the corresponding system \cite{RefF1,RefF2,RefF3,RefF4,RefF5,RefF6,RefF7}. It measures disorder in the phase space trajectories by defining a Shannon based entropy \cite{RefF3,RefF4,RefF5,RefF6,RefF7}. Several entropy measures have already been proposed to quantify the dynamical complexity \cite{RefF3,RefF4,RefF5,RefF6,RefF7}. Among the all, recurrence based entropy measure is more effective that can be applied in any dimensional system. Further, a weighted recurrence entropy measure has been proposed in \cite{RefF8} which is more robust than recurrence based entropy \cite{RefF9}.\par
This paper is organized as follows: In section 2, a noise induced blood flow system is described. The noise is taken as power noise with $\beta \in [0,1]$. Changes in dynamics are discussed by measuring the periodicity. Further, chaos has been quantified by $0-1$ test. Section 3 discusses the asymptotic behaviour of the dynamics by phase space analysis. Based on weighted recurrence, the disorder in the phase space has been investigated. The corresponding disorders have also been quantified by entropy analysis. A strong correlation between the dynamics and complexity is verified. Finally, a conclusion is given in section 4.     
\section{Complex dynamics of a noise induced blood flow}
\label{sec:1}
\subsection{Noise induced blood flows system}
\label{sec:11}
\begin{figure}[h!]
\resizebox{1.00\columnwidth}{!}{%
\includegraphics{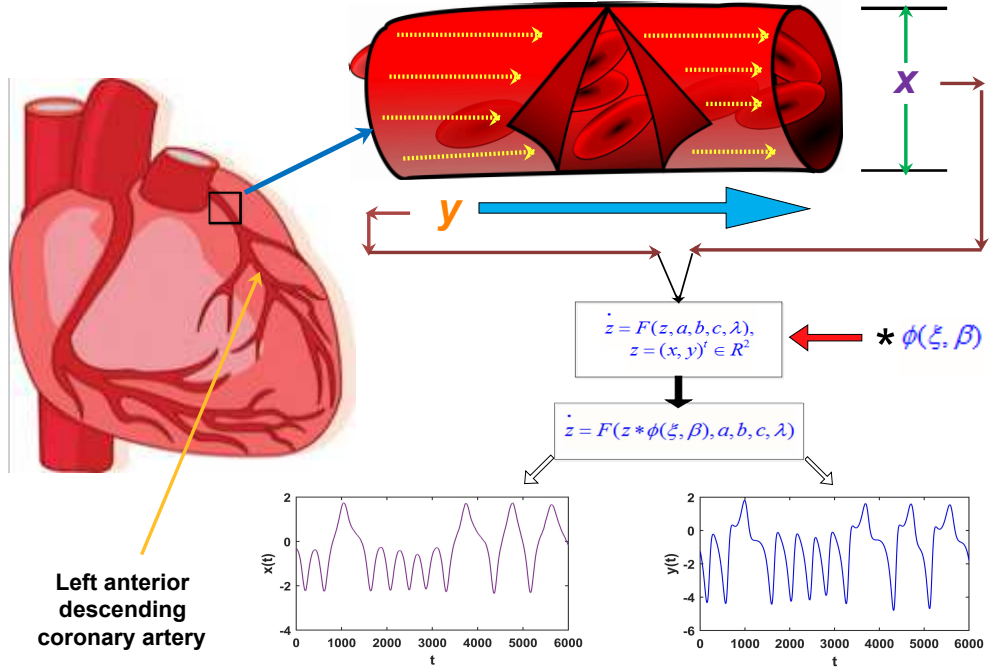} }
\caption{Schematic diagram for a blood flow phenomenon. At the top of the diagram red colored tube represents a portion of a blood vessel of a left anterior descending coronary artery in the human heart. The variable $x$ and $y$ indicates diameter of the vessel and pressure of the flowing blood in the vessel respectively. $\dot{z}=F(z,a,b,c, \lambda)$ represents blood flow system containing $z=(x,y)^t \in R^2$ as variable and $a,b,c, \lambda$ as parameters. The `$\ast$' on the top of the $\leftarrow$ indicates incorporation of multiplicative noise $\phi(\xi,\beta)$ with $\dot{z}=F(z,a,b,c, \lambda)$. The new system is then represented by $\dot{z}=F(z,a*\phi(\xi,\beta),b,c, \lambda)$. At the bottom of the diagram, left (violet) and right (blue) graph represents solution components $x(t)$ and $y(t)$ of $\dot{z}=F(z,a*\phi(\xi,\beta),b,c, \lambda)$. In order to compute $x$ and $y$ over an time interval, we have taken $\beta=0.5,a=0.21, b=0.15, c =-0.15, \lambda=-0.169$.}
\label{fig:schem}       
\end{figure}

We have introduced a noise $\phi(\xi,\beta)$ to the blood flow model \cite{RefA7} by
\begin{equation} \label{eq:Eq1}
(\dot{x},\dot{y})=(a*\phi(\xi,\beta)-bx-cy,-\lambda (1+b)x-\lambda (1+c)y+\lambda x^{3}),
\end{equation}
where $\phi(\xi,\beta)$ represents samples of $\frac{1}{f^{\beta}}$-noise ($f$ being the frequency) and $a,b,c, \lambda$ represents system parameters. The underlying mechanism are given in a schematic diagram (see Fig.\ref{fig:schem}). In Fig.\ref{fig:schem}, the zoomed portion of an artery showing the blood flow in a vessel. The `yellow' arrows represent the direction of the flow. We consider $x$ and $y$ as variables that indicates respective changes in diameter of the vessel and pressure of the fluid (containing blood cells) over time. The system $\dot{z}=F(z,a*\phi(\xi,\beta),b,c, \lambda)$, $z=(x,y) \in R^2$ represent same as given in (\ref{eq:Eq1}), where $z^T$ represents transpose of $z$ and $a*\phi(\xi,\beta)$ indicates product of $a$ and $\phi(\xi,\beta)$. The function $\phi(\xi,\beta)$ indicates noise component $\xi$ whose power spectrum $S(f) \sim f^{\beta}$. 
\subsection{Noise induced Periodicity}
\label{sec:12}
\begin{figure}[h]
\resizebox{1.00\columnwidth}{!}{%
\includegraphics{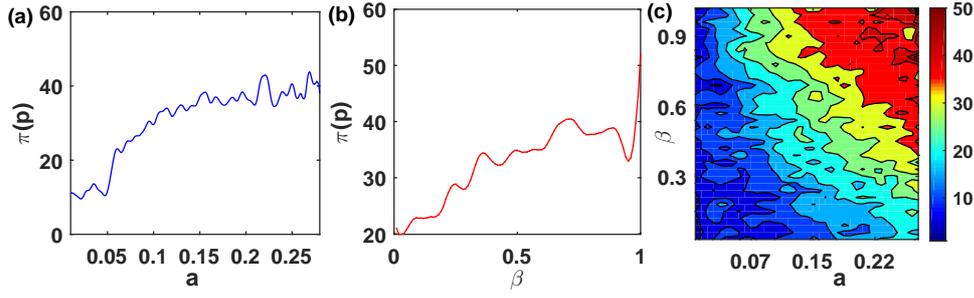} }
\caption{(a) and (b) represents $\pi(p)$ vs. $a \in [0.01,0.28]$ (with fixed $\beta=0.5$) and $\beta \in [0,1]$ (with fixed $a=0.21$) plots respectively. (c) represents  contour of $\pi(p)$ over $(a,\beta) \in [0.01,0.28] \times [0,1]$. The color bar indicates values of $\pi(p)$. In each case, $\pi(p)$ are  computed by counting the number of periods $p$ of $x$ solution component of (\ref{eq:Eq1}) with the initial condition $x(0)=0.1,y(0)=0.2$ and $b=0.15, c =-0.15, \lambda=-0.169$.}
\label{fig:1}       
\end{figure}

In this section, we investigate periodic behaviour of the system (\ref{eq:Eq1}) for $a \in [0.01,0.28]$ (with fixed $\beta=0.5$) and $\beta \in [0,1]$ (with fixed $a=0.21$) respectively. To do this, we have calculated periodicity $\pi(p)$ for one of the solution component of (\ref{eq:Eq1}). The corresponding fluctuations are given in Fig.\ref{fig:1}a and b respectively.    
From both Fig.\ref{fig:1}a and b, an increasing trend in the respective fluctuation of $\pi(p)$ can be observed. It implies that the multi-periodicity increases over $a \in [0.01,0.28]$ (with fixed $\beta=0.5$) and $\beta \in [0,1]$ (with fixed $a=0.21$). It indicates changes in complex dynamics of (\ref{eq:Eq1}) with the increasing $a$ and $\beta$ respectively. However, a non-increasing pattern can be seen in Fig.\ref{fig:1}a, within a small range of $a \in [0.01,0.05]$. It signifies small changes in the corresponding dynamics. Moreover, a certain change in the gradient of $\pi(p)$ curves observed from both the figures indicate fast changes in the dynamics of (\ref{eq:Eq1}). So, $\pi(p)$ can describe the variation in the dynamics with the changes of $a \in [0.01,0.28]$ and $\beta \in [0,1]$ respectively. We further investigated periodicity $\pi(p)$ of (\ref{eq:Eq1}) for $(a,\beta) \in [0.01,0.28] \times [0,1]$. The corresponding contour is given in Fig.\ref{fig:1}c. From the figure, the increasing trend in the fluctuation of $\pi(p)$ corresponds the change in the dynamics of (\ref{eq:Eq1}). Moreover, certain changes in the colors of some regions with a small duration can also be observed in Fig.\ref{fig:1}c. It indicates the faster change in $\pi(p)$ compare to same on the other regions. \par
As changes in $\pi(p)$ with varying $a$ and $\beta$ do not recognize the nature of the dynamics, a measure is thus needed to quantify the chaotic and non-chaotic behaviour of (\ref{eq:Eq1}). In the following section, both nature has been investigated using $0$-$1$ chaos test:
\subsection{Chaos test}
\label{sec:13}
In $0$-$1$ method, only one solution $\{x_k\}_{k=1}^{N}$ ($N$ being the length of the component) of a system is considered. Then it decomposed into into two components $p_{k}^{\eta}$ and $q_{k}^{\eta}$ using the transformations
\begin{equation} \label{eq:Eq2}
p_{k}^{\eta}=\sum_{j=1}^{k}x(j) \cos(j\eta), \quad q_{k}^{\eta}=\sum_{j=1}^{k}x(j) \sin(j\eta), 
\end{equation}
where $\eta \in (0,\pi)$. For a non-chaotic system, $(p_{k}^{\eta},q_{k}^{\eta})$-plot reveals a regular geometric structure. On the other hand, Brownian motion like structure can be observed in the corresponding $(p_{k}^{\eta},q_{k}^{\eta})$-plot for a chaotic system. \par
\begin{figure}[h!]
\resizebox{1.00\columnwidth}{!}{%
\includegraphics{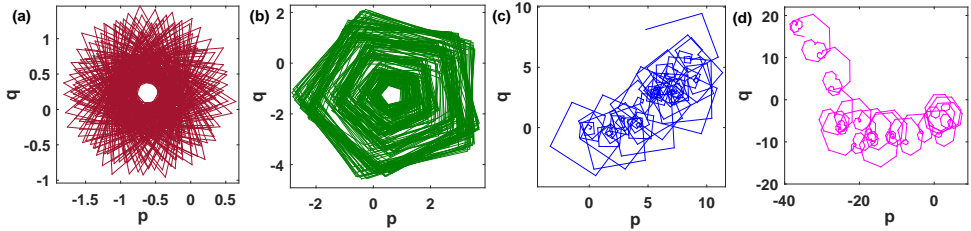} }
\caption{(a)-(d) represents $p$ vs. $q$ plots with $(a, \beta)=(0.01,0.2)$, $(a, \beta)=(0.21,0.2)$, $(a, \beta)=(0.21,0.5)$ and $(a, \beta)=(0.21,1)$ respectively. The values of $b, c , \lambda$ are taken as $0.15, -0.15, -0.169$ respectively.}
\label{fig:2}       
\end{figure}
\begin{figure}[h]
\begin{center}
\resizebox{1.00\columnwidth}{!}{%
\includegraphics{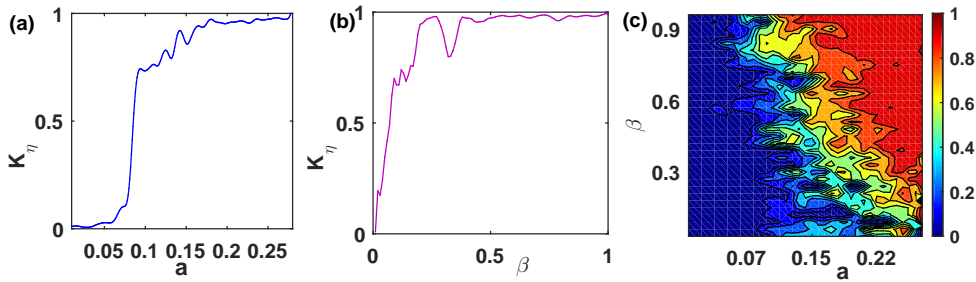} }
\caption{(a) and (b) represents $K_\eta$ vs. $a \in [0.01,0.28]$ (with fixed $\beta=0.5$) and $\beta \in [0,1]$ (with fixed $a=0.21$) plots respectively. (c) represents  contour of $K_\eta$ over $(a,\beta) \in [0.01,0.28] \times [0,1]$. The associate color bar indicates the values of $K_\eta$. In each cases, $K_\eta$s have been computed by taking $k_{cut}=\frac{N}{10}$, $N$ being the length of the $x$-component. For the numerical simulation, we have taken $N=10000$.}
\label{fig:3}       
\end{center}
\end{figure}

For our purpose, $x$-solution of (\ref{eq:Eq1}) is considered and we have used the notation $(p,q)$ in place of $(p_{k}^{\eta},q_{k}^{\eta})$. Fig.\ref{fig:2} shows $(p,q)$-plots for some fixed $a$ and $\beta$ respectively. In Fig.\ref{fig:2}a and b, regular geometrical structure can be seen in the respective $(p,q)$-plots. It indicates that the corresponding dynamics are non-chaotic at $(a, \beta)=(0.01,0.2)$ and $(a, \beta)=(0.21,0.2)$ respectively. From Fig.\ref{fig:2}c and d, Brownian motion like structure can be observed in the respective $(p,q)$-plot. It corresponds chaotic dynamics of the system at the respective parameters $(a, \beta)=(0.21,0.5)$ and $(a, \beta)=(0.21,1)$. In this way,  regular as well as the chaotic dynamics of (\ref{eq:Eq1}) can be characterized for the entire ranges of $a$ and $\beta$ respectively. \vskip 3pt
In \cite{RefE2}, it has been seen that the diffusive and the non-diffusive behaviour of a system can also be measured from the above $(p,q)$-plots using $M_{k}^{\eta}$, where 
\begin{equation} \label{eq:Eq7}
M_{k}^{\eta}=\lim_{N \to \infty}\frac{1}{N}\sum_{j=1}^{N}[p_{j+k}^{\eta}-p_{j}^{\eta}]^2+[q_{j+k}^{\eta}-q_{j}^{\eta}]^2.
\end{equation}
Here, $k \leq k_{cut}$ and $k_{cut}<<N$. In practice, the value of $k_{cut}$ is taken by $k_{cut}=\frac{N}{10}$. \vskip 3pt
Moreover, it can be also observed in \cite{RefE2} that, the asymptotic growth of $M_{k}^{\eta}$ can quantifies both chaotic and non-chaotic dynamics of a system. The asymptotic growth ($K_\eta$) is defined by
\begin{equation} \label{eq:Eq8}
K_\eta=\lim_{k\to\infty} \frac{\log {M_{k}^{\eta}}}{\log k}.
\end{equation}
The values of $K_\eta \approx 1$ and $0$ indicates chaotic and regular dynamics of the system \cite{RefE2}. \vskip 3pt
To quantify both chaotic and non-chaotic dynamics of (\ref{eq:Eq1}), we have studied the fluctuation of $K_\eta$ over $a \in [0.01,0.28], \beta=0.5$ and $\beta \in [0,1], a=0.21$ respectively. The corresponding $K_\eta$ vs. $a$ and $\beta$ graphs are shown in Fig.\ref{fig:3}a and b respectively. From Fig.\ref{fig:3}a, it can be seen that $K_\eta \approx 1$ when $a \geq 0.18$. It indicates chaotic dynamics of (\ref{eq:Eq1}) for $a \geq 0.18, \beta=0.5$. As $K_\eta \not \approx 1$ for $a \in[0.01,0.18), \beta=0.5$, it corresponds regular dynamics of (\ref{eq:Eq1}). 
On the other hand, fluctuation in Fig.\ref{fig:3}a shows that values of $K_\eta \approx 1$ for almost all $\beta \in [0,1]$, except $\beta \in (0.15,0.2)$. It implies chaotic and non-chaotic dynamics of (\ref{eq:Eq1}) for all $\beta \in [0,1]-(0.15,0.2)$.
Further, fluctuation of $K_\eta$ over $(a,\beta) \in [0.01,0.28] \times [0,1]$ has also been investigated. Fig.\ref{fig:3}c shows the corresponding contours. In Fig.\ref{fig:3}c, the red regions correspond $K_\eta \approx 1$. It indicates chaotic dynamics of the system (\ref{eq:Eq1}). On the other hand, blue, green and yellow regions correspond $K_\eta \not \approx 1$ and hence non-chaotic dynamics of (\ref{eq:Eq1}) respectively. The sharp boundary (black) indicates abrupt changes in the fluctuation of $K_\eta$ that signifies fast changes in the dynamics of (\ref{eq:Eq1}). In this way, regular as well as chaotic dynamics have been quantified using $0$-$1$ method.  \par 

In the next section, we have studied dynamical complexity of the system (\ref{eq:Eq1}) using phase space based weighted entropy.
\section{Dynamical complexity of noise induced blood flow}
In order to study the dynamical complexity, we first investigate asymptotic dynamics of (\ref{eq:Eq1}) under the variation of parameters $a$ and power $\beta$ respectively. 
\label{sec:2}
\subsection{Phase space analysis}
\label{sec:21}
\begin{figure}[h]
\begin{center}
\resizebox{1.00\columnwidth}{!}{%
\includegraphics{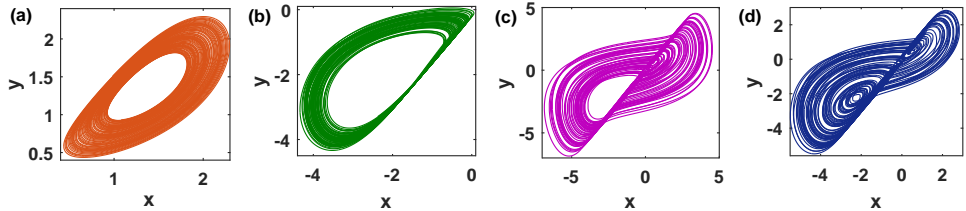} }
\caption{(a) and (b) represents 2D projection of the phase spaces with $(a, \beta)=(0.01,0.2)$, $(a, \beta)=(0.21,0.2)$ respectively. For $(a, \beta)=(0.21,0.5)$ and $(a, \beta)=(0.28,1)$, the respective 2D phase portraits are given in (c) and (d). In order to draw the phase spaces, we have solved the system (\ref{eq:Eq1}) with $x(0)=0.1, y(0)=0.2$ and $a=0.15, b=-0.15, \lambda=-0.169$.}
\label{fig:4}       
\end{center}
\end{figure}

To investigate the asymptotic behaviour of (\ref{eq:Eq1}), we study the nature of the corresponding phase spaces of (\ref{eq:Eq1}) over $a \in [0.01,0.28]$, $\beta \in [0,1]$ respectively. Some of the corresponding 2D phase spaces are shown in Fig.\ref{fig:4}. Fig.\ref{fig:4} a and b show regular phase spaces containing multiperiodic orbits with $ (a, \beta)=(0.01,0.2)$ and $(a, \beta)=(0.21,0.2)$ respectively. On the other hand, irregular trajectory movements can be observed in both the Fig.\ref{fig:4} c and d with the respective $(a, \beta)=(0.21,0.5)$ and $(a, \beta)=(0.28,1)$. It implies that, the respective phase spaces are chaotic and also indicates a correlation with the results of $0-1$ tests. The same can be investigated for the entire range of $a$ and $\beta$. Now, phase space can reflects long-term dynamics of a system. It indicates that the dynamical complexity of (\ref{eq:Eq1}) can be quantified by measuring a disorder in the phase space.

In the next, we investigate both disorder and dynamical complexity of (\ref{eq:Eq1}) using weighted recurrence and its entropy.
\subsection{Disorder and weighted entropy analysis}
\label{sec:22}
\begin{figure}[h]
\begin{center}
\resizebox{1.00\columnwidth}{!}{%
\includegraphics{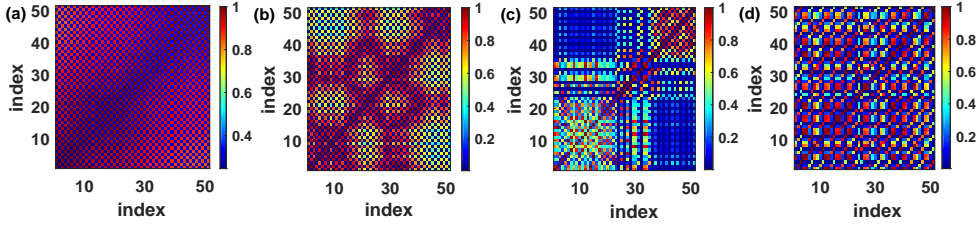} }
\caption{(a) and (b) represents weighted matrix plots with $(a, \beta)=(0.01,0.2)$, $(a, \beta)=(0.21,0.2)$ respectively. For $(a, \beta)=(0.21,0.5)$ and $(a, \beta)=(0.28,1)$, the same are given in (c) and (d) respectively. In order to construct the $(\omega_{ij})_{50 \times 50}$, we have considered the system (\ref{eq:Eq1}) with $x(0)=0.1, y(0)=0.2$ and $a=0.15, b=-0.15, \lambda=-0.169$. The corresponding color bars indicate values of weights $\omega_{ij}$.}
\label{fig:5}       
\end{center}
\end{figure}

For an $n$-dimensional phase space $P=\{(\overrightarrow{x}_i): (\overrightarrow{x}_i) \in R^n\}$, $i=1,2,..,N$ ($N$ being length of a trajectory), weighted recurrence can be defined by a matrix $W=(\omega_{ij})_{n \times n}$, where $\omega_{ij}=e^{-\|\overrightarrow{x}_i-\overrightarrow{x}_j\|}$. The weights $\omega_{ij}$ correspond exponential divergence of the point $\overrightarrow{x}_j$ from $\overrightarrow{x}_i$ and vice verse. As the movements of the trajectory is related to the distance between $\overrightarrow{x}_j$ and $\overrightarrow{x}_i$, the disorder in the phase space can be described by the weights $\omega_{ij}$. In fact, all types of disorder in $P$ can be described by $W$. We now investigate disorder in the phase space of (\ref{eq:Eq1}) using weighted recurrence $\omega_{ij}$. Fig.\ref{fig:5}a and b shows weighted matrix plots with $(a, \beta)=(0.01,0.2)$, $(a, \beta)=(0.21,0.2)$ respectively. From the figures, it can be observed that variation in colours of the corresponding weighted recurrence plots are very small. It indicates small disorderedness in the respective phase spaces. On the other hand, various structures can be seen in both the Fig.\ref{fig:5}c and d. It indicates highly disordered structure in the corresponding phase spaces. As disorder in the phase spaces is strongly correlated with the corresponding weighted recurrence, so this analysis indicates that the complexity of the system (\ref{eq:Eq1}) can be quantified using weighted recurrence based entropy \cite{RefF9}.   

\begin{figure}[h]
\begin{center}
\resizebox{1.00\columnwidth}{!}{%
\includegraphics{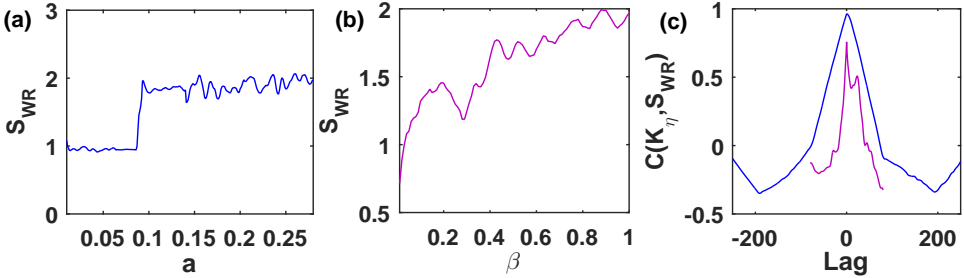} }
\caption{(a) and (b) represents $S_{WR}$ vs. $a$ and $\beta$ with $a \in [0.01,0.28]$ (with fixed $\beta=0.5$) and $\beta \in [0,1]$ (with fixed $a=0.21$) respectively. To compute $S_{WR}$, the probabilities are counted with 20 bins in each case. (c) represents cross-correlation curves $C(K_\eta,S_{WR})$ for $a \in [0.01,0.28], \beta=0.5$ (blue) and $\beta \in [0,1], a=0.21$ (red) respectively. On computing $C(K_\eta,S_{WR})$, the ranges of respective lags are taken as $[-200,200], [-100,100]$.}
\label{fig:6}       
\end{center}
\end{figure}

For a an weighted matrix $(\omega_{ij})_{N \times N}$, the weighted entropy ($S_{WR}$) is defined by
\begin{equation}
S_{WR}=-\sum_{s_k\in S}^{}p(s_k)\log p(s_k),
\end{equation}
where $p(s_k)$ is the probability of $s_k$ and $S$ is a collection $\{s_k: s_k=\frac{1}{N}\sum_{j=1}^{N}\omega_{kj}, 1 \leq k \leq N\}$. In practice, probability is computed by bit counting method. We have computed fluctuation of $S_{WR}$ over $a \in [0.01,0.28]$ (with fixed $\beta=0.5$) and $\beta \in [0,1]$ (with fixed $a=0.21$) respectively. The corresponding graphs are shown in Fig.\ref{fig:6}a and b respectively. 

From the figures, it can be observed that both the fluctuation having similar trend as that have been observed in Fig.\ref{fig:3}a and b. To verify the correlation between the fluctuations of $K_\eta$ and $S_{WR}$, we have done a cross-correlation analysis for both $a$ and $\beta$. The respective cross-correlation curves are given in Fig.\ref{fig:6}c. From Fig.\ref{fig:6}c, it can be observed that maximum values of both $C(K_\eta,S_{WR})$ are always greater than $0.85$. It indicates strong correlation between $S_{WR}$ and $K_\eta$ for $a \in [0.01,0.28]$ (with fixed $\beta=0.5$) and $\beta \in [0,1]$ (with fixed $a=0.21$) respectively. \par
\begin{figure}[h]
\begin{center}
\resizebox{1.00\columnwidth}{!}{%
\includegraphics{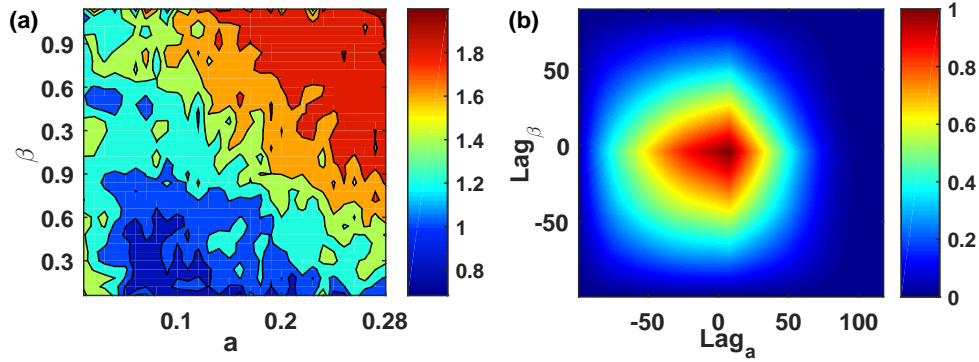} }
\caption{(a) represents $S_{WR}$ vs. $(a,\beta) \in [0.01,0.28] \times [0,1]$. To compute $S_{WR}$, the probabilities are counted with 20 bins in each case. (b) represents $2D$ cross-correlation contour with $(Lag_a,Lag_\beta) \in [-50,50] \times [-100,100]$.}
\label{fig:7}       
\end{center}
\end{figure}

We further investigate effect of $(a,\beta) \in [0.01,0.28] \times [0,1]$ on $S_{WR}$. The corresponding contour is given in Fig.\ref{fig:7}a. From both the Fig.\ref{fig:7}a and Fig.\ref{fig:3}c, it can be observed that the fluctuation of $S_{WR}$ and $K_\eta$ are almost similar. It indicates the existence of a correlation between $S_{WR}(a,\beta)$ and $K_\eta(a,\beta)$. To confirm this, we have done a cross-correlation analysis. Fig.\ref{fig:7}b shows the corresponding $2D$ cross-correlation contour. It can be seen from the contour that the maximum value $0.98$ corresponds at $(Lag_a,Lag_\beta)=(0,0)$, where $Lag_a$ and $Lag_\beta$ represents $Lag$ of $a$ and $\beta$ respectively. It assures that  $S_{WR} \sim K_\eta$ for $(a,\beta) \in [0.01,0.28] \times [0,1]$.  
\section{Conclusion}
\label{sec:3}
In this article, we have studied the effect of power noise on a blood flow system. The effects have been investigated by periodicity, clustering between chaotic and non chaotic dynamics and complexity. In the periodic analysis, increasing trends in $\pi(p)$ have been observed over $a \in [0.01,0.28]$ (with fixed $\beta=0.5$), $\beta \in [0,1]$ (with fixed $a=0.21$) and $(a,\beta) \in [0.01,0.28] \times [0,1]$ respectively. It indicates a change in dynamics or more precisely increasing instability in (\ref{eq:Eq1}). Moreover, fast as well as slow changes in the dynamics have been also identified. To recognize regular and chaotic regime, we have applied $0-1$ test. In this analysis, chaotic as well as non chaotic windows have been recognized using the respective fluctuation of $K_\eta$ over same $a$ (with fixed $\beta=0.5$), $\beta$ (with fixed $a=0.21$) and $(a,\beta)$. In the next, a correlation between the $K_\eta$ and asymptotic dynamics have been observed by applying phase space analysis. To quantify this, the corresponding disorder was defined using weighted recurrence concept. It has been seen that weighted recurrence can successfully recognize the disorder in the phase space. The respective entropy fluctuations $S_{WR}(a)$, $S_{WR}(\beta)$ and $S_{WR}(a,\beta)$ show a strong correlation with the same of $K_\eta$. The correlations have been confirmed by standard cross-correlation analysis. So, the whole analysis indicates that the noise induced blood flow system (2) possesses chaos within a certain range of system parameter and power of the noise. Moreover, entropy analysis assures that complexity of the system increases with increasing $a \in [0.01,0.28],$ $\beta \in [0,1]$.

\end{document}